\documentclass[11pt]{article}

\usepackage{graphicx}
\usepackage{amsmath, amsthm, amsfonts}
\usepackage[numbers]{natbib}
\usepackage{mathtools}

\usepackage[a4paper, total={6in, 8in}]{geometry}
\usepackage{caption}
\usepackage{subcaption}

\begin{document}

\title{Inherent  Negative Refraction on Acoustic Branch of Two Dimensional Phononic Crystals}
\author{Sia Nemat-Nasser\\
Department of Mechanical and Aerospace Engineering\\
University of California, San Diego\\
La Jolla, CA, 92093-0416 USA\\
sia@ucsd.edu}
\maketitle

\begin{abstract}

Guided by theoretical predictions, we have demonstrated experimentally the existence of negative refraction on the two lowest  \textit{acoustic-branch} passbands (shear and longitudinal modes) of a simple two-dimensional phononic crystal consisting of an \textit{isotropic stiff} (aluminum) matrix and square-patterned \textit{isotropic compliant} (PMMA) circular inclusions.
\textit{At frequencies and wave vectors where the refraction on the acoustic-branch passbands is negative, the effective mass-density and the effective stiffness tensors of the crystal are positive-definite, and this is an inherent property of such phononic crystals.}
The equi-frequency contours and energy flux vectors as fuctions of the phase-vector components, reveal a rich body of refractive properties that can be exploited to realize, for example, beam splitting, focusing, and frequency filtration on the lowest passbands of the crystal where the dissipation is minimum. By proper selection of material and geometric parameters these phenomena can be realized at remarkably low frequencies (large wave lengths) using rather small simple two-phase unit cells. 

\end{abstract}

\newpage

\textit{Introduction}:
Photonic composites consisting of a periodic arrangement of split-ring resonators and wires display negative refraction on their optical branch and possess frequency-dependent simultaneously negative effective dielectric permittivity, $\epsilon_{eff}(\omega)$, and magnetic permeability, $\mu_{eff}(\omega)$, \cite{Smith2000,shelby2001experimental}.  Composites of this kind have become known as "left-handed" or "meta-materials".

Similarly periodic elastic composites consisting of periodically distributed elastic inclusions in an elastic matrix
exhibit band structure \cite{minagawa1976harmonic}, negative refraction \cite{nemat2015refraction}, and effective properties that are frequency-dependent \cite{nemat2017unified}.  Early experimental demonstrations of negative refraction and focusing properties of phononic crystals have employed water as the matrix material, within which stiff inclusions are periodically distributed, revealing negative refraction on the second (optical branch) passband \cite{yang2004focusing,ke2005negative, sukhovich2008negative}. 
Phononic crystals of this kind can  transmit dilatational waves only, since their matrix material lacks shear resistance.  Hence their acoustic branch contains only one (dilatational) passband involving positive refraction only.
 It is known that photonic and phononic crystals with anisotropic constituents can display negative refraction on their \textit{acoustic branch} \cite{notomi2000theory,nemat2017unified}, but this has not been demonstrated experimentally.  
We point out that, using numerical simulations, negative refraction has been shown \cite{Luo2002all} to be possible on a narrow region of the acoustic branch of a photonic crystal consisting of circular holes placed in a square-lattice pattern within a homogeneous and isotropic dielectric matrix.  By interfacing the crystal with air along its (11)-direction, it is shown theoretically that negative refraction can occur in a narrow frequency at the far edges of the first passband.

In this paper we show theoretically and demonstrate experimentally that a two-dimensional phononic crystal consisting of an  \textit{isotropic stiff} (aluminum) matrix and square-patterned embedded \textit{isotropic compliant} (PMMA) circular inclusions \textit{does display negative refraction over a broad region on its acoustic branch for both shear and longitudinal modes.} 
Interestingly, our theoretical predictions and preliminary experimental results show that if a compliant (PMMA) material is used for the matrix (instead of aluminum) and stiff (aluminum) material for inclusions, then only the longitudinal mode of the acoustic branch would display negative refraction. We have followed this by a series of band-structure and associated energy-flux calculations using various material properties for the matrix and inclusions.  It turns out that, for negative refraction to occur on the first (shear mode) passband, the stiffness of the matrix material must be considerably greater than that of the inclusions. The existence of large Poisson's ratios does enlarge the region of negative refraction in the phase-space, but does not appear to be essential for the existence of such region. Similarly, the values of the mass-density do influence this phenomenon but only quantitatively.  As should be expected, for very small Poisson's ratios, the results obtained using plane stress or plane strain would be essentially the same, but not for large values of the Poisson ratios. In either case negative refraction is shown to be an integral part of the acoustic branch of phononic crystals. 
The presence of negative and positive refractions on the acoustic-branch passbands allows one to focus, filter, and split multifrequency beams at very low frequencies with many potential applications.

\textit{Sample Properties}:
The sample is a simple doubly periodic elastic composite composed of $10~ mm$ square aluminum unit cells, each containing a central $4.46 ~mm$ diameter circular PMMA inclusion. The material properties are:

$$\noindent E_1^M=E_2^M=68 ~GPa,~~\nu^M=0.33,~~\rho^M=2700 ~kg/m^3;$$
$$\noindent E_1^I=E_2^I=3.0~ GPa,~~\nu^I=0.40,~~\rho^I=1200 ~kg/m^3,
$$
where $E_j^M$ and $E_j^I$, $j=1,2$,  are the elastic moduli of the matrix and inclusions respectively, with the corresponding Poisson ratios and mass densities being denoted by $\nu^M, \nu^I$ and $\rho^M, \rho^I$.

Here the focus is on the acoustic branch which, in two dimensional elastic phononics, includes two passbands, one (the lowest passband)  is  shear mode and the other is  longitudinal mode. For the material properties given above, the equi-frequency contours accompanied by the energy flux vectors,
the variation of the  shear- and longitudinal-mode frequency along  the $\Gamma, X, M, \Gamma$ lines, and the frequency surfaces of the acoustic-branch passbands are  shown in Figure \ref{TR}
for plane stress.  
As is seen from the directions of the energy flux vectors in Figures \ref{TR}(a,b) and suggested by the peak values of the frequency in Figures \ref{TR}(c,d), 
both shear and longitudinal modes display positive and negative refraction; note the peak values of frequency, one between $X$ and $M$ for the shear mode, and two for the longitudinal mode: one between $\Gamma$ and X and the other between X and M. These theoretical results have been used to guide the sample design and subsequent experimental measurements.
Also, we have included the three-dimensional Figure \ref{TR}(c) to show that the first two passbands are distinct and do not intersect one another.
We note that similar results are obtained for the plane strain case, but it is the plane stress estimates which most closely follow the experimental data.
\begin{figure}\label{TR}
\centering
\begin{subfigure}[t]{0.45\linewidth}
\includegraphics[scale=0.4, trim=0cm 0cm 0cm 0cm, clip=true]{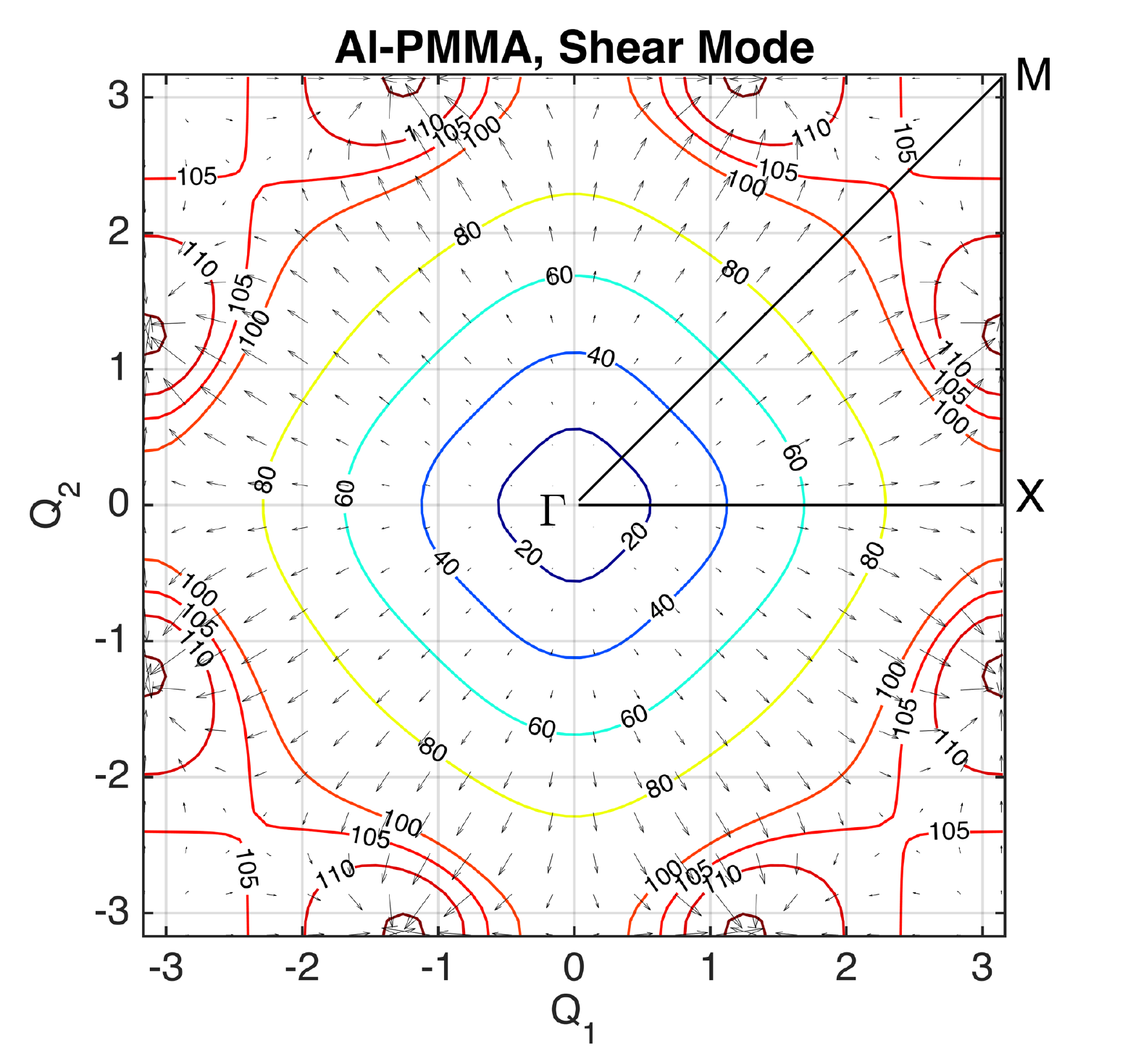}
\caption{}\label{fig:1a}	
\end{subfigure}
\begin{subfigure}[t]{0.45\linewidth}
\includegraphics[scale=0.4, trim=0cm 0cm 0cm 0cm, clip=true]{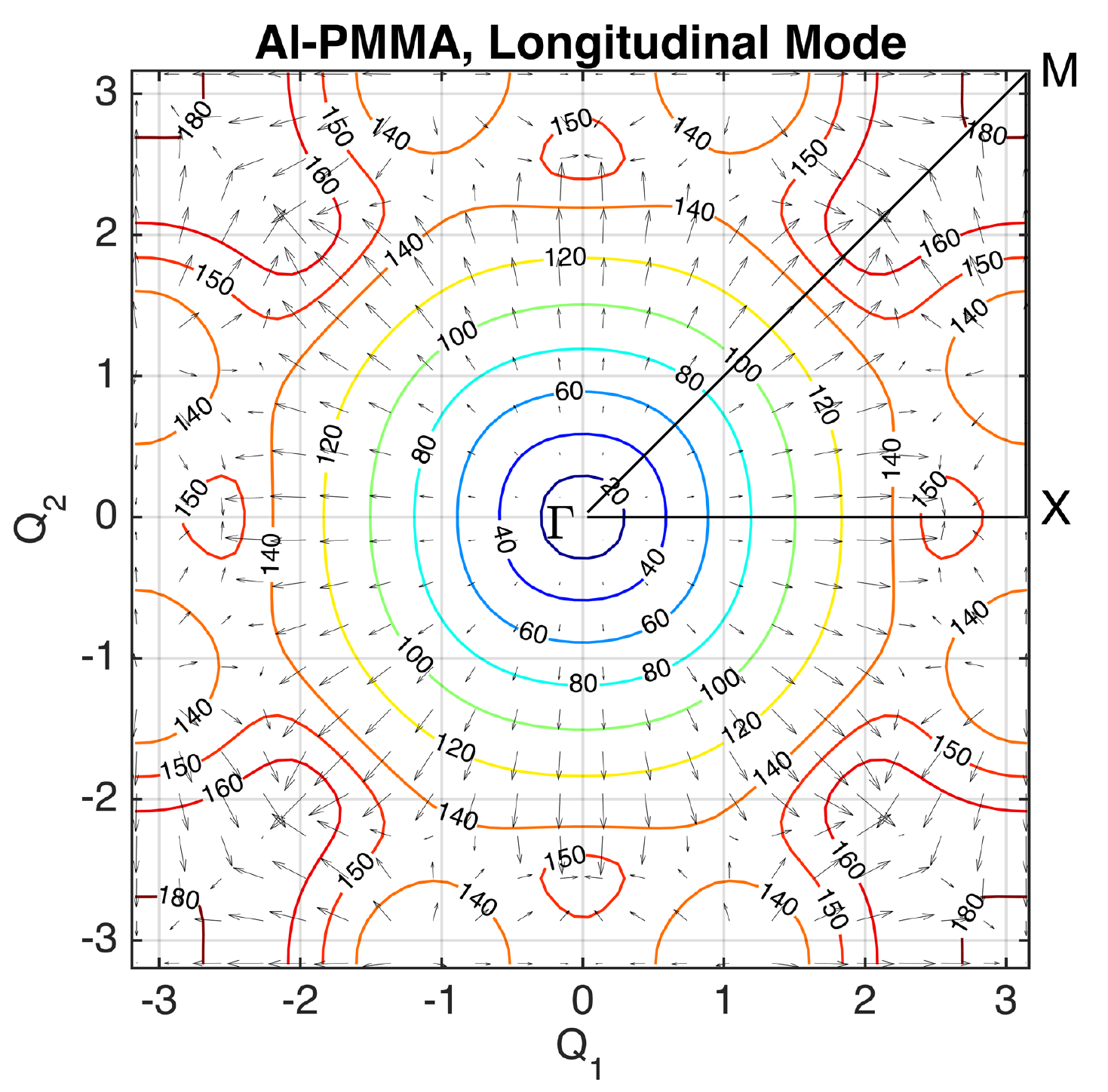}
\caption{}\label{fig:1b}	
\end{subfigure}
\begin{subfigure}[t]{0.45\linewidth}
\includegraphics[scale=0.40, trim=0cm 0cm 0cm 0cm, clip=true]{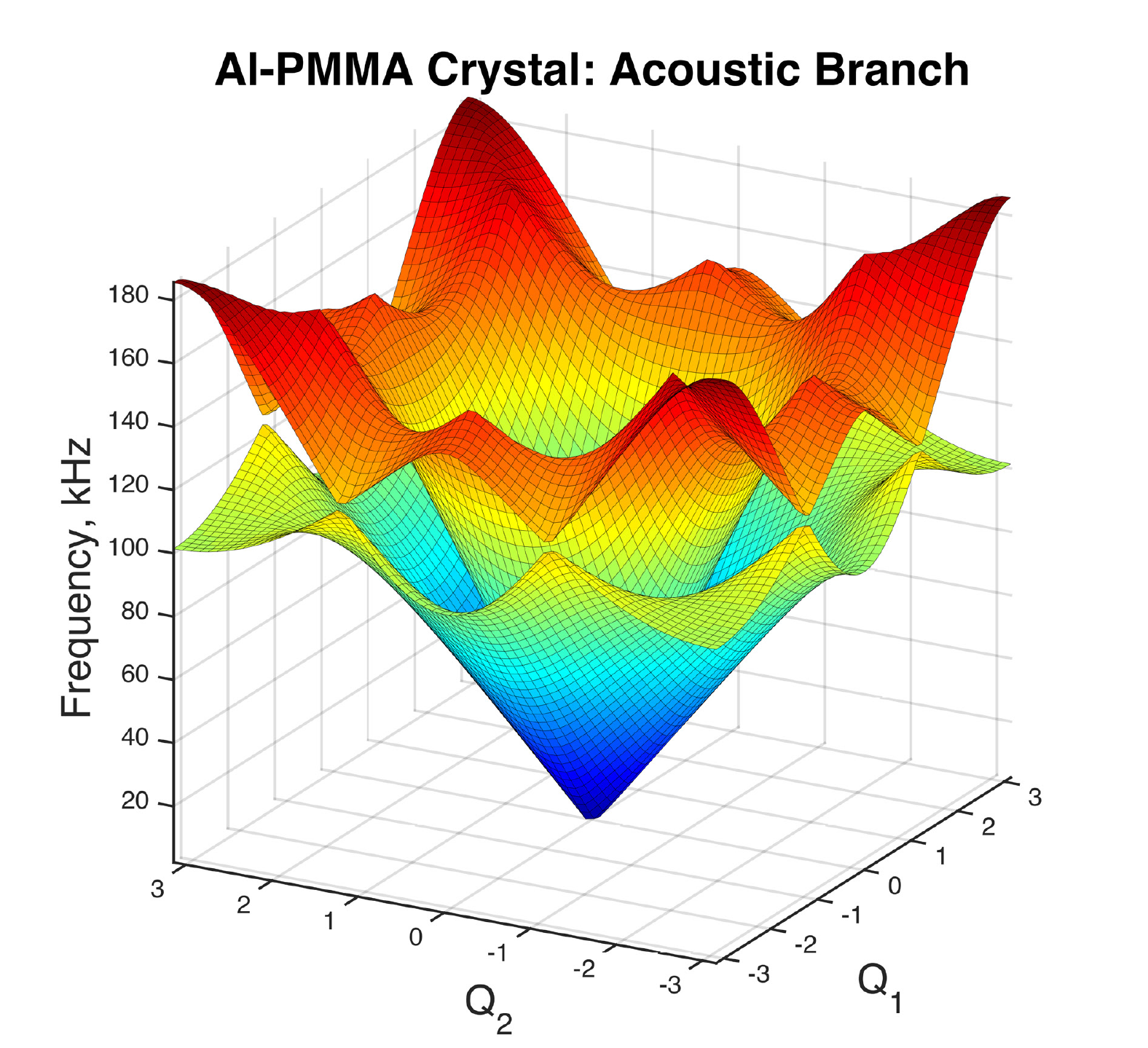}
\caption{}\label{fig:1c}	
\end{subfigure}
\begin{subfigure}[t]{0.45\linewidth}
\includegraphics[scale=0.40, trim=0cm 0cm 0cm 0cm, clip=true]{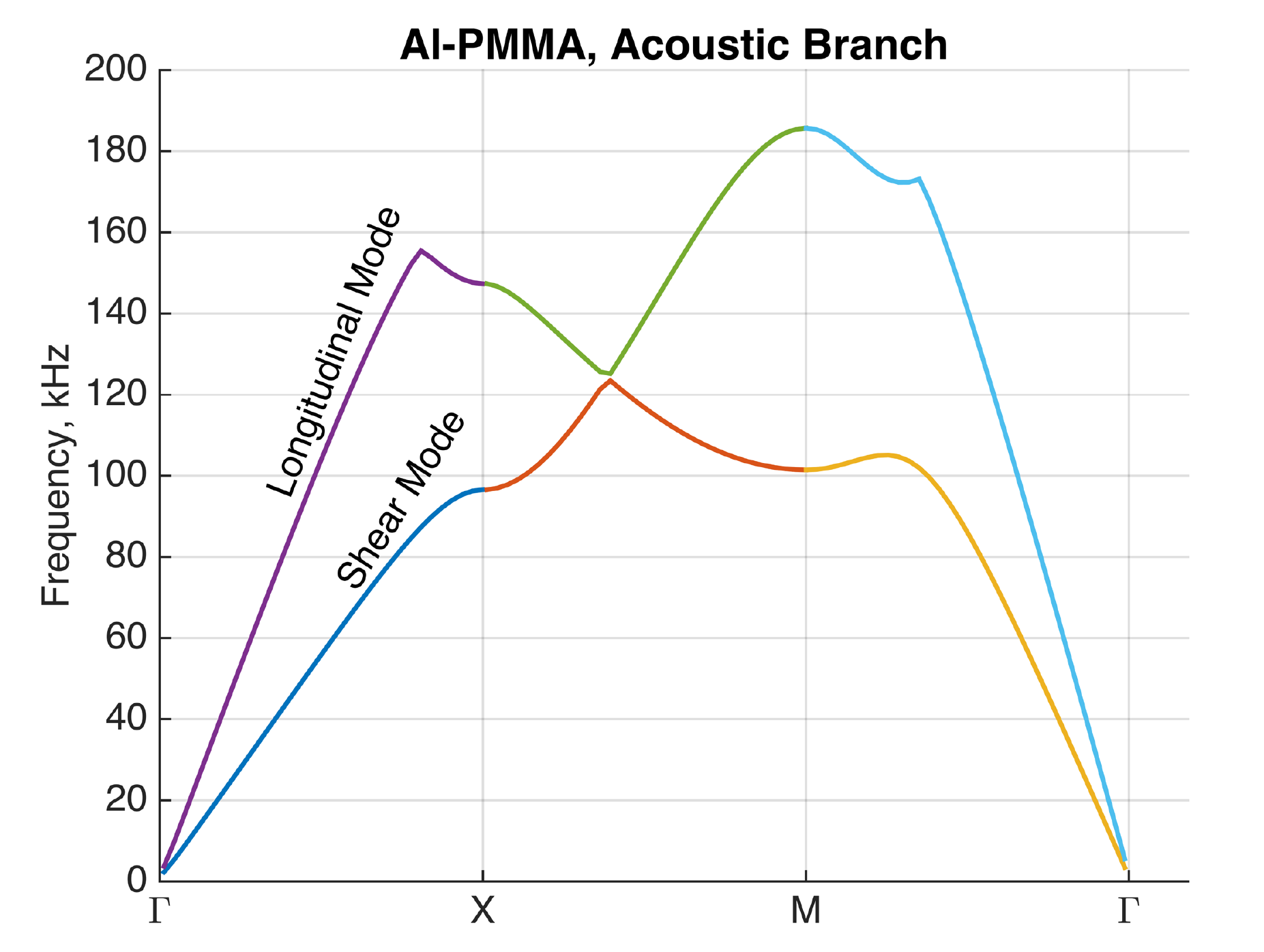}
\caption{}\label{fig:1d}	
\end{subfigure}
\caption{Acoustic-branch equi-frequency contours and energy flux vectors of: (a) shear mode and, (b), longitudinal mode passbands.  (c) Frequency surfaces. (d) Frequency variation along the $\Gamma, X, M, \Gamma$ lines for shear and longitudinal modes.}
\label{TR}
\end{figure}
\begin{figure}
\centering
\begin{minipage}[b]{0.45\linewidth}
\hspace{-1.85cm}
\includegraphics[scale=0.40, trim=0cm 0cm 0cm 0cm, clip=true]{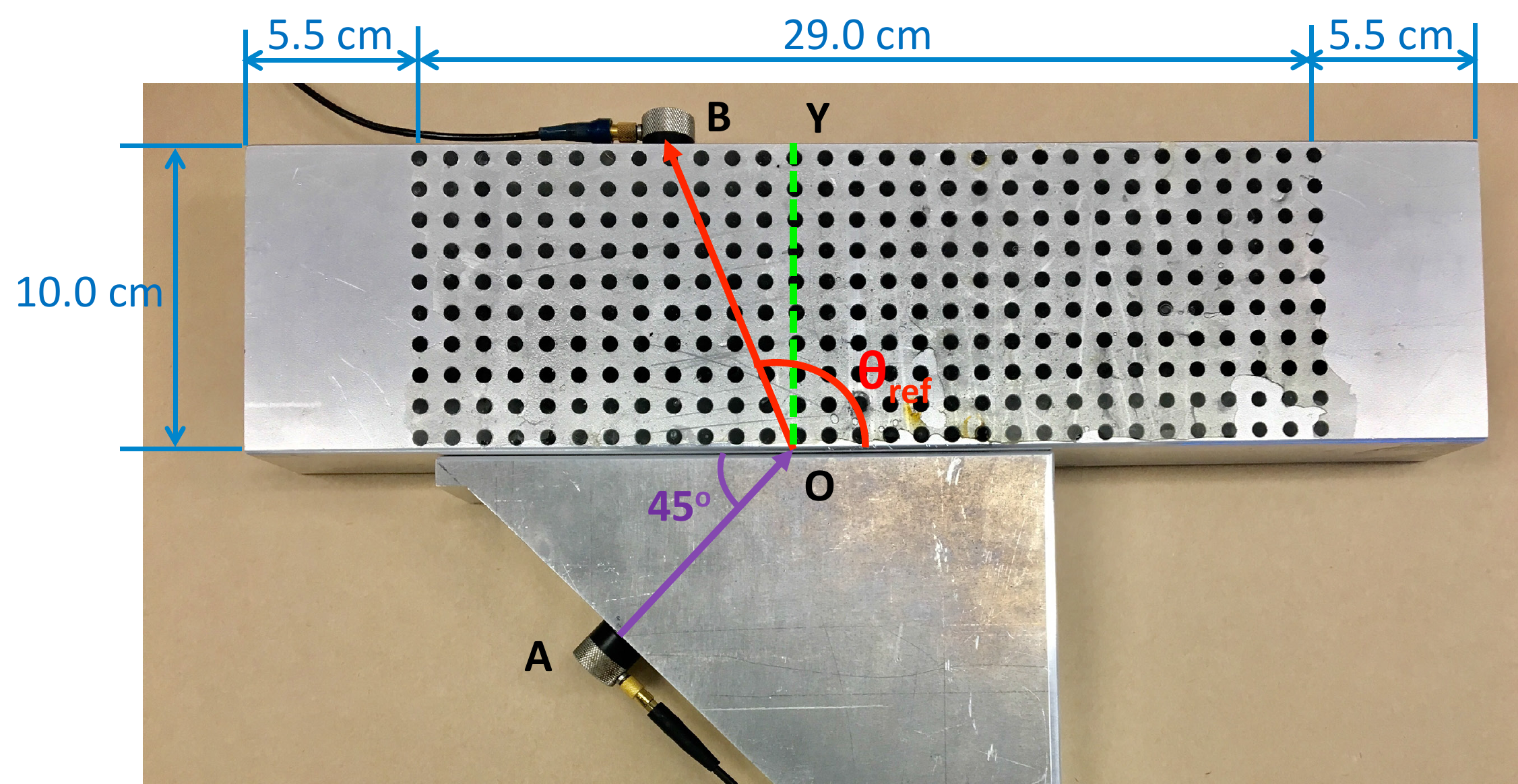}
\end{minipage}
\begin{minipage}[b]{0.40\linewidth}
\hspace{0.5cm}
\includegraphics[scale=0.32, trim=1cm 0cm 0cm 0cm, clip=true]{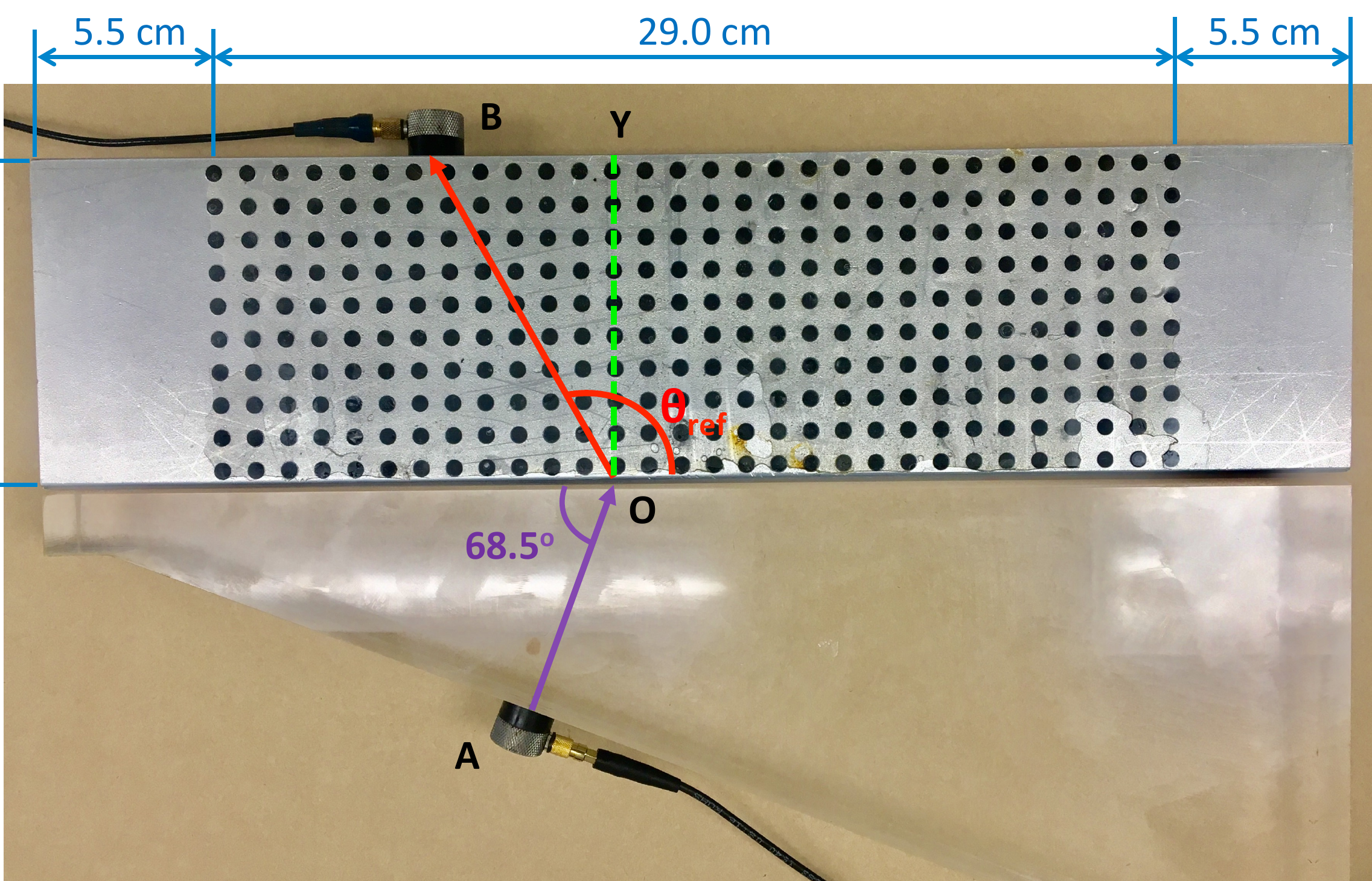}
\end{minipage}
\caption{An aluminum sample containing 10 mm spaced, 4.76 mm diameter, circular PMMA inclusions with:  (Left) a $45^o$ homogeneous  aluminum wedge, and (Right) a $68.5^o$ PMMA wedge.  Transducer A sends an incident signal through the homogeneous  wedge. The signal is received by transducer B  and is recorded.}
\label{ExpSetup}
\end{figure}

\textit{Experimental Setup and Results}:
The experimental setups for shear- and longitudinal-signal measurements are shown in Figures \ref{ExpSetup}. The homogeneous wedge for the shear-mode is  $45^o$ aluminum and that for the longitudinal-mode is  $68.5^o$ PMMA. These are chosen such that each could support the corresponding incident wave within the desired frequency range.

\begin{figure}
\centering
\begin{minipage}[b]{0.45\linewidth}
\includegraphics[scale=0.30, trim=1.5cm 0cm 0cm 0cm, clip=true]{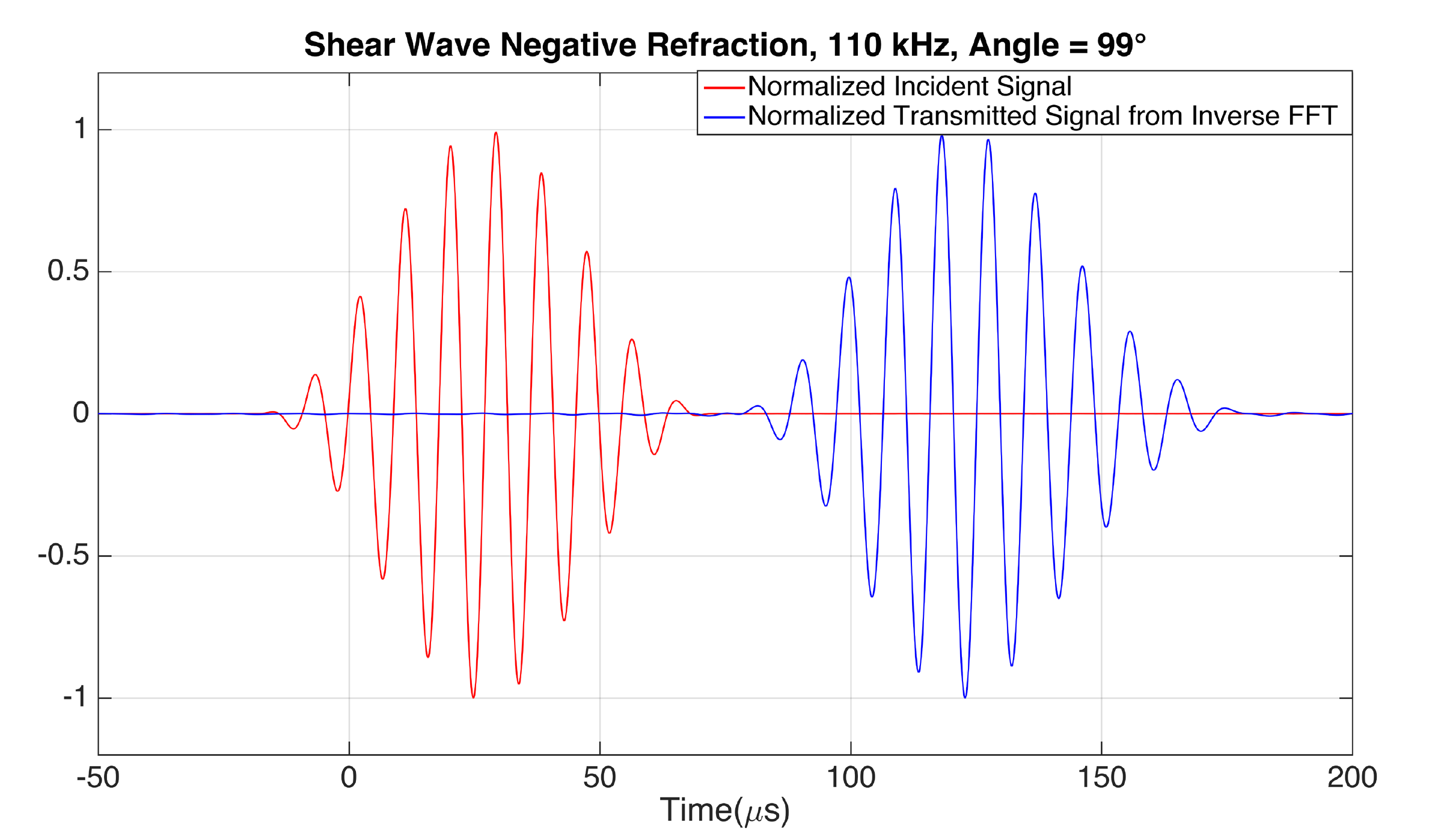}
\end{minipage}
\begin{minipage}[b]{0.45\linewidth}
\includegraphics[scale=0.30, trim=0cm 0cm 0cm 0cm, clip=true]{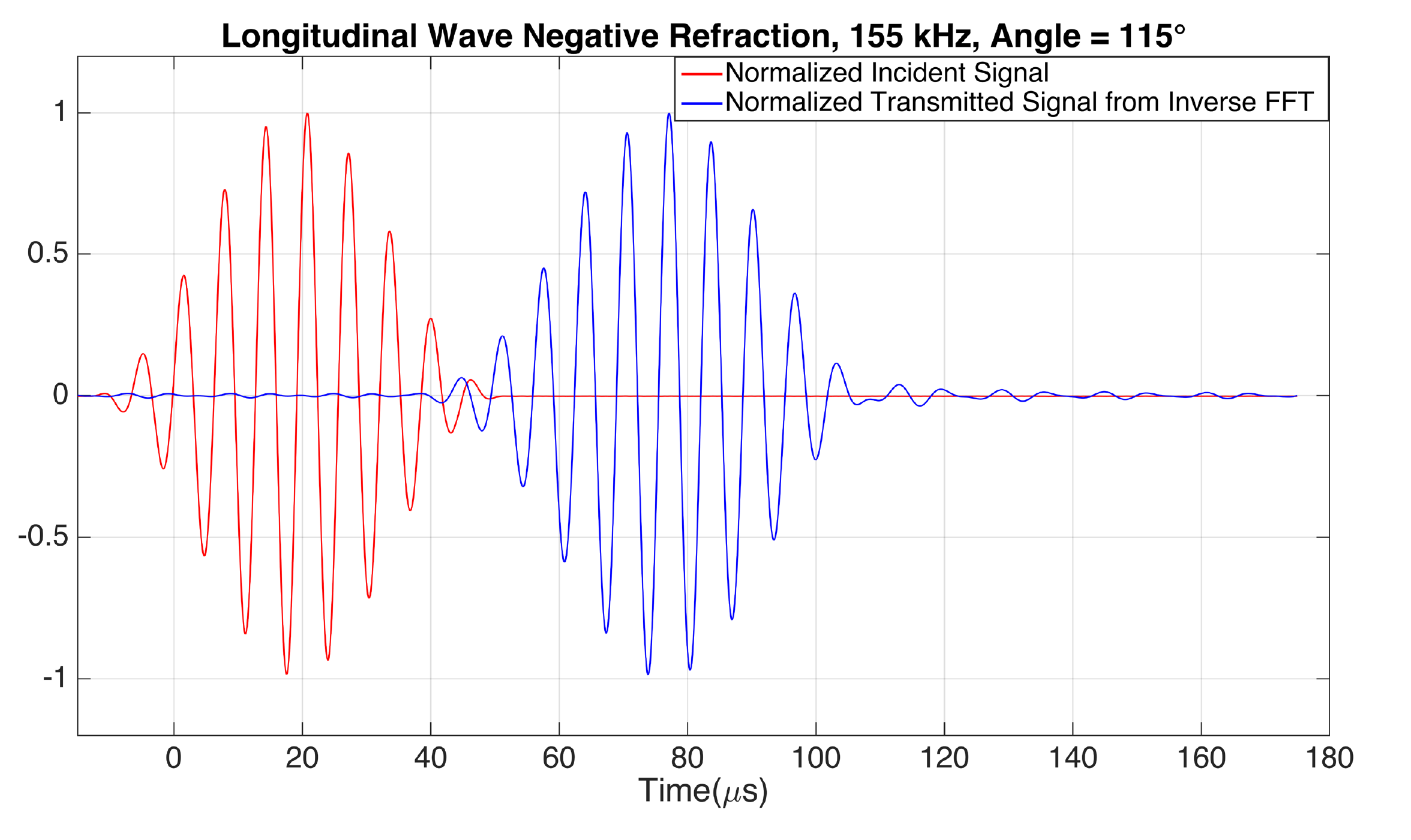}
\end{minipage}
\caption{Typical input-output signals; negative refraction, (Left) shear mode, and (Right) longitudinal mode.}
\label{expSL}
\end{figure}

The incident signal is transmitted by transducer A through the homogeneous wedge and is received by traducer B and recorded. The refraction angle is manually measured. Its value depends on the incident frequency.  Typical input-output signals are shown in Figure \ref{expSL}.  FFT is used to process the raw output signals and filter out surface waves and other accompanying signals on the tail of the actual signal. The actual signals are dominant and clearly distinguishable, especially for negative refraction.  

\begin{figure}
\begin{center}
\includegraphics[scale=0.80, trim=0cm 0cm 0cm 0cm, clip=true]{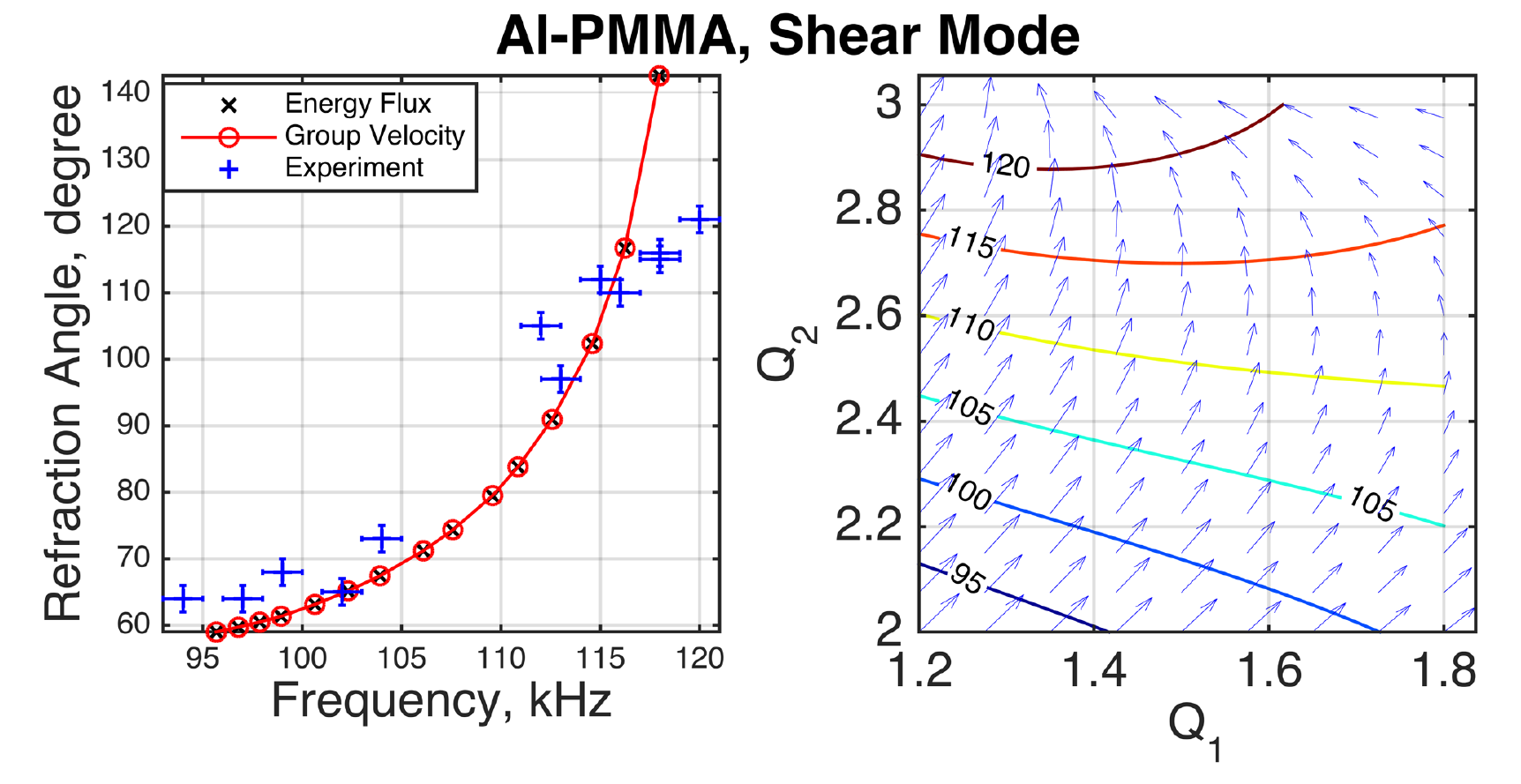}
\caption{(Left): Group velocity and energy flux directions and experimental results. (Right): Equi-frequency contours and energy flux vectors of the first (shear mode) passband; $Q_1=k_1a$ and $Q_2=k_2a$ are the normalized $x_1$ and $x_2$ components of the wave vector.}
\label{expS1}
\end{center}
\end{figure}

\begin{figure}
\begin{center}
\includegraphics[scale=0.80, trim=0cm 0cm 0cm 0cm, clip=true]{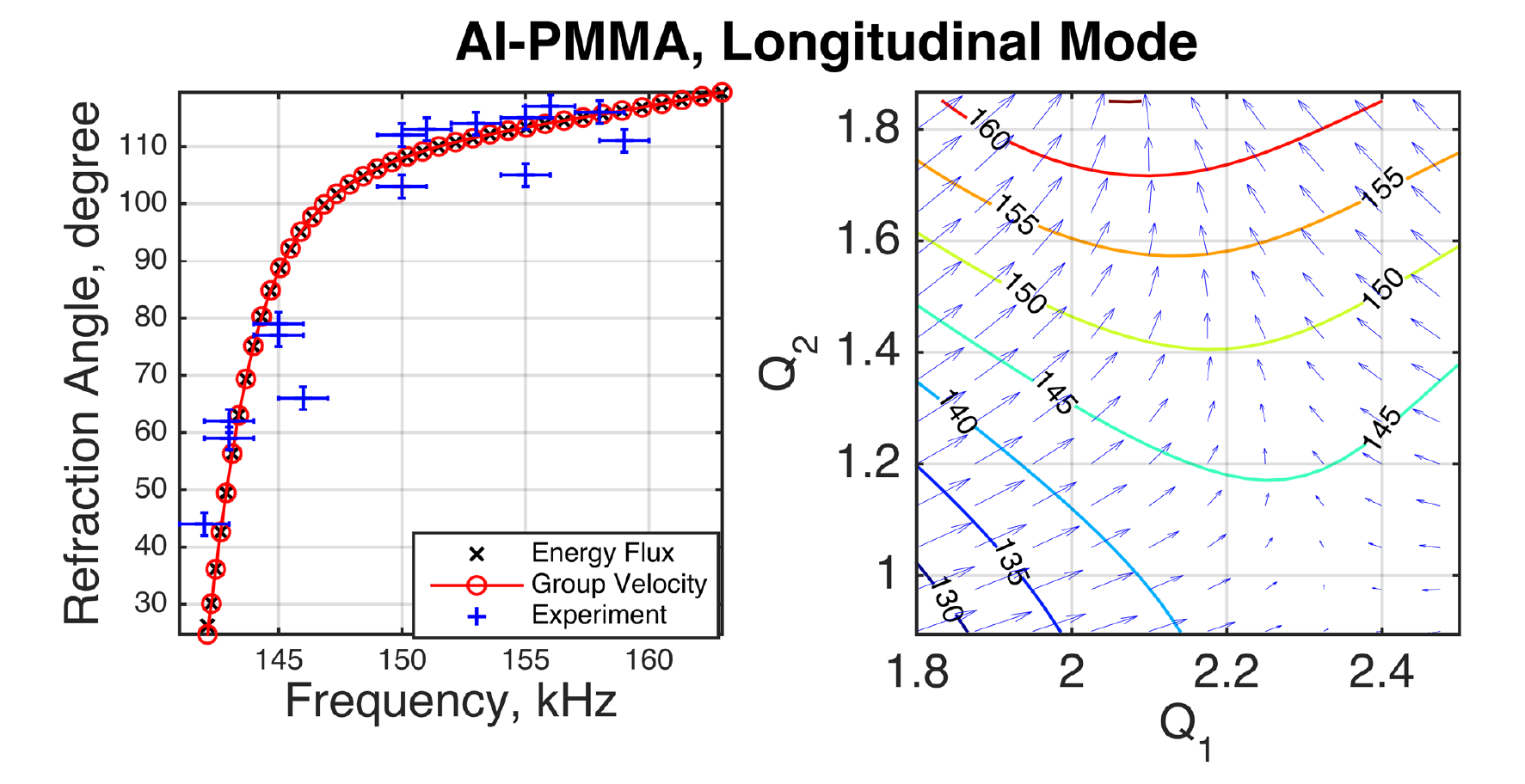}
\caption{(Left): Group velocity and energy flux directions for $Q_1=2.25$, and experimental results. (Right): Equi-frequency contours and energy flux vectors of the first (shear mode) passband; $Q_1=k_1a$ and $Q_2=k_2a$ are the normalized $x_1$ and $x_2$ components of the wave vector.}
\label{expL1}
\end{center}
\end{figure}

The measurements are guided by the theoretical results shown in Figure \ref{expS1}(Right) for the shear mode and in Figure \ref{expL1}(Right) for the longitudinal mode.  The corresponding energy-flux and group velocity directions are shown in Figures \ref{expS1}(Left) and \ref{expL1}(Left), respectively, together with the experimental results. As is seen, the refraction directions estimated from the direction of energy flux are identical with those obtained from the group-velocity directions.

The refraction angle for each frequency is manually measured.  It turns out that the results are not sensitive to small variations in either frequency or the location where the refraction is being measured.  The error bars in the x- and y-directions in Figures \ref{expS1}(Left) and \ref{expL1}(Left) reflect this fact.
The experimental results closely follow the predictions based on the plane stress, giving credence to the basic theoretical approach,  summarized in the last section of this paper.  

In the homogeneous wedge of fixed angle $\theta_0$, the $x_1$-phase component is given by $k_1=\frac{coa(\theta_0)\omega}{c_0a_1}$, where $\omega$ is the  frequency  and $c_0$ is the corresponding wave-speed in the wedge.  This fact is reflected in the calculation of the group-velocity and energy-flux directions for the shear mode but, for the longitudinal mode, an average $Q_1=k_1a_1=2.25$ is used. 
For the shear mode, the plane-stress and plane-strain cases yield similar results, but not for the longitudinal mode. It turns out that the plane stress results closely follow the experimental data as discussed above.

\begin{figure}[htbp]
\begin{center}
\begin{minipage}[b]{0.45\linewidth}
\includegraphics[scale=0.40, trim=0.cm 0.cm 0.cm 0.cm, clip=true]{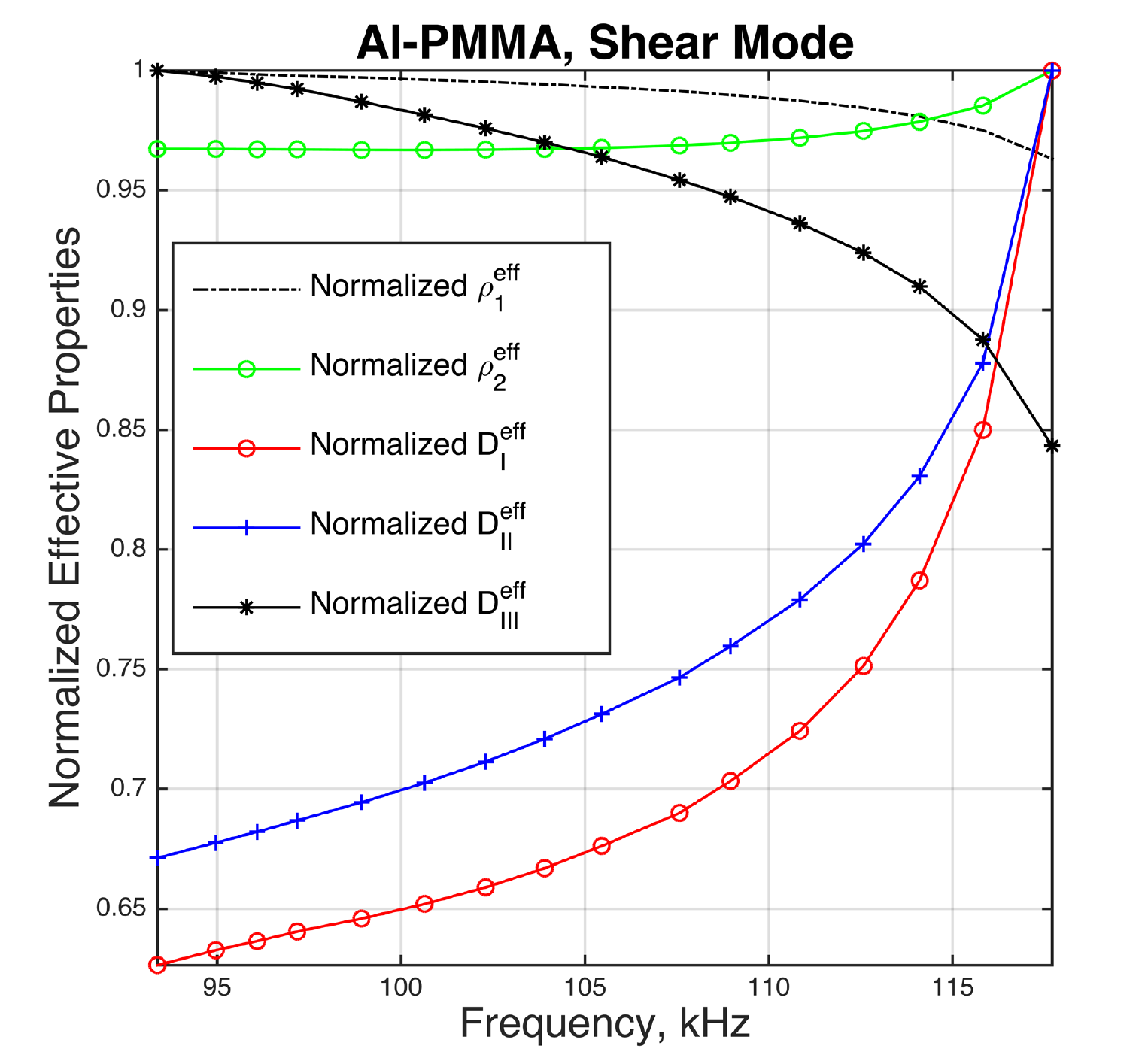}
\end{minipage}
\begin{minipage}[b]{0.45\linewidth}
\includegraphics[scale=0.40, trim=0.cm 0.cm 0.cm 0.cm, clip=true]{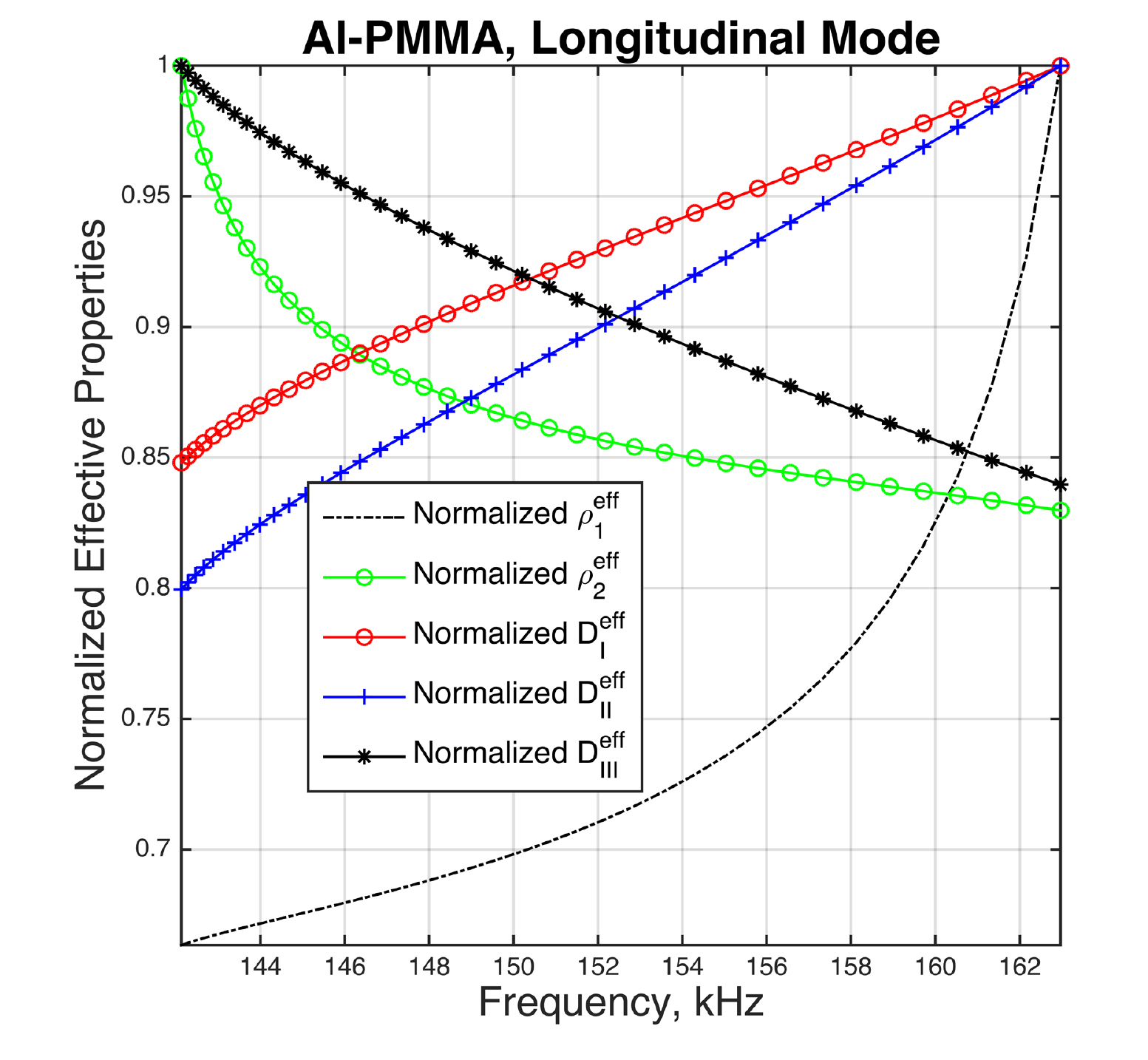}
\end{minipage}
\caption{Normalized effective mass-densities and the principal (eigen) values of the effective compliance matrix as functions of frequency. For clarity in the graphical display, elements of each array are normalized with respect to the corresponding largest element.}
\label{Effprop}
\end{center}
\end{figure}

Remarkably,  the effective compliance (inverse of stiffness)  and the effective mass-density tensors are both positive-definite for positive and negative refraction angles, as shown in Figure \ref{Effprop}.  
In two dimensional elasticity, the components of the effective compliance tensor, $D_{ijkl}^{eff}$, can be represented by a
$3 \times 3$ matrix and the mass-density by a $2 \times 2$ diagonal matrix. For the  effective compliance tensor to be positive-definite, its eigenvalues, $D_I^{eff},D_{II}^{eff}, $ and $D_{III}^{eff}$, must all be positive. 
In Figure \ref{Effprop} we have presented the normalized effective mass-densities and the normalized eigenvalues of the effective compliance matrix as functions of the frequency, where, for normalization, the elements of each array is divided by its own largest (positive) element. As is seen these effective properties are strictly positive over the entire considered frequency range corresponding to the results shown in Figures \ref{expS1}(Left) and \ref{expL1}(Left). 
% %%%%%%%%%%%%%%%%%%%%%%%%%%

At first this may appear surprising since a counterpart does not exist for two-dimensional photonic crystals, nor for phononic crystals in anti-plane shearing, where negative refraction on the acoustic branch can be realized only if we use an anisotropic matrix material \cite{notomi2000theory,nemat2017unified}.
% &&&&&&&&&&&&&&&&&&&&&&&&&&&&&&&&&&&&&&&&&&
The EM properties of an isotropic material are completely defined by an electric permittivity, $\epsilon$, and a magnetic permeability, $\mu$.  
In two dimensional isotropic elasticity, on the other hand, in addition to the mass-density $\rho$, there are  two elastic constants, the shear modulus $\mu$, and Poisson's ratio $\nu$, forming a $3\times3$ matrix, 
\begin{eqnarray}\label{compliance}
[D]=\frac{1}{\mu}
\left[ \begin{array}{ccc}
\frac{\kappa+1}{8}&\frac{\kappa-3}{8}&0\\
\frac{\kappa-3}{8}&\frac{\kappa+1}{8}&0\\
0&0&\frac{1}{2}
\end{array} \right],~~~~\kappa=
\left[ \begin{array}{c}
3-4\nu~~~plane~strain\\
\frac{3-\nu}{1+\nu}~~~plane~stress
\end{array} \right],
\end{eqnarray}
that relates the strain components,
$[\epsilon]=\{\epsilon_{11}~~\epsilon_{22}~~\epsilon_{12}=\epsilon_{21}\}^T$, 
 to the stress components,
$[\sigma]=\{\sigma_{11}~~\sigma_{22}~~\sigma_{12}=\sigma_{21}\}^T$,
by
$[\epsilon]=[D][\sigma]$.
The principal (eigen) vectors of $[D]$ are the rows of 
$[V]=[\frac{-1}{\sqrt2}~~\frac{1}{\sqrt2}~~0;0~~0~~1;\frac{1}{\sqrt2}~~\frac{1}{\sqrt2}~~0]$ so that 
$[D^\prime]=[V][D][V]^T$ is a diagonal matrix with diagonal elements
$D_I=D_{II}=\frac{1}{2\mu}$ and
%$D_{III}=\frac{\kappa-1}{4\mu}$. 
$D_{III}=\frac{(1-2\nu)}{2\mu}$ for plane strain, and
$D_{III}=\frac{1-\nu}{2\mu(1+\nu)}=\frac{1-\nu}{E}$ for plane stress. 
In the transformed space, the strain and stress components are 
$[\epsilon^\prime]=[V][\epsilon]=
[\frac{\epsilon_{22}-\epsilon_{11}}{\sqrt2}~~ \epsilon_{12}~~\frac{\epsilon_{22}+\epsilon_{11}}{\sqrt2}]$, and 
$[\sigma^\prime]=[V][\sigma]=
[\frac{\sigma_{22}-\sigma_{11}}{\sqrt2}~~\sigma_{12}~~\frac{\sigma_{22}+\sigma_{11}}{\sqrt2}]$, which are related by 
$[\epsilon^\prime]=[D^\prime][\sigma^\prime]$.
The first two components of $[\epsilon^\prime]$ correspond to the shearing and the last one to the longitudinal deformation modes.
For small value of the Poisson ratios, the value of $D_{III}$ approaches $\frac{1}{2\mu}$, yet the associated strain and stress components retain their longitudinal characteristics. 

The periodically embedded inclusions in such an isotropic elastic slab results in in-plane wave-dispersion and produces band structure for Bloch-form periodic waves.   The corresponding  
acoustic branch will always  include a shear and a longitudinal passband, both of which can display negative refraction.  
In contrast, the acoustic branch of a two-dimensional phononic crystals in anti-plane shear (SH) mode, or photonic crystals in transverse electric (TE) or transverse magnetic (TM) mode, has only one passband with only positive refraction if the constituents of the crystal are isotropic.  
 
The fact that an incident plane shear or longitudinal wave can be refracted at positive or negative angles depending on the frequency, allows for splitting a multifrequency plane waves. For example, 
%%%%
consider two plane shear-waves that are transmitted through a homogeneous aluminum wedge at a $\theta_0=45^o$ incident angle with respect to its interface with the composite, one at frequency $\omega_1=105 kHz$ and the other at frequency
$\omega_2=115 kHz$. Figure \ref{expS1}(Left) shows that the first wave will travel in the crystal at a positive angle of about $70^o$, whereas the second one at an angle of about $110^o$ relative to the positive $x_1$-axis. 
Also, all plane waves of frequencies greater that 
$112^o$ would have negative refraction while those of frequencies less than $112^o$ will have positive refraction.
In addition, if two plane waves of  $115 kHz$ common frequency are transmitted through the crystal a distance, say $2d$ apart, one at a $+45^o$ angle and the other at a $-45^o$ angle, they would focus at a distance $L=d/tan(20^o) $ from the interface.

\textit{Field Equations and Periodic Solution}:
Considered is a doubly periodic elastic composite composed of rectangular unit cells of dimensions $a_1$, $a_2$. A typical unit cell, $\Omega$, includes a concentric inclusions, $\Omega_1$.   
For Bloch-form time-harmonic waves of frequency $\omega$, the field variables are all proportional to $e^{-i\omega t}$ which can be incorporated into the field equations,
\begin{equation}\label{field1}
\sigma_{lj,l}+\omega^2\rho u_j=0;\quad
\frac{1}{2}(u_{j,l}+u_{l,j})-D_{ljmn}\sigma_{mn}=0,\quad
 j,l,m,n=1,2,
\end{equation}
where comma followed by an index denotes differentiation with respect to the corresponding coordinate,  $\rho$ is the mass density, $D_{ljmn}=D_{mnlj}$ is the elastic compliance  tensor, and
\begin{eqnarray}\label{field2}
\left[ \begin{array}{c}
u_j\\
\sigma_{lj}\\
\end{array} \right]=
\left[ \begin{array}{c}
u_j^p(x_1,x_2,x_3)\\
\sigma_{jl}^p(x_1,x_2,x_3)\\
\end{array} \right]e^{i(k_1x_1+k_2x_2+k_3x)}.
\end{eqnarray}
Here the superimposed $p$ stands for the periodic part.
The geometry, the mass-density, and the elastic compliance
are periodic with the periodicity of the unit cell. 
For both plane strain and plane stress problems, the nonzero components of the compliance tensor, $D$, 
\begin{equation}\label{p4}%11
D_{1111}=D_{11},\quad
D_{1122}=D_{12}=D_{21},\quad
D_{2222}=D_{22},\quad
D_{1212}=D_{2121}=D_{33},
\end{equation}
are defined in equations \ref{compliance}.

To calculate the frequency passbands of the crystal, subject to the Bloch periodicity condition, express the periodic part of the field variables as follows:
\begin{eqnarray}\label{field6}
\left[ \begin{array}{c}
u_j^p\\
\sigma_{lj}^p\\
\end{array} \right]=\sum_{n_1,n_2=-N}^{+N}
\left[ \begin{array}{c}
U_j^{n_1,n_2}\\
S_{jl}^{n_1,n_2}\\
\end{array} \right]
e^{i2\pi({\frac{n_1x_1}{a_1}+\frac{n_2x_2}{a_2})}}.
\end{eqnarray}

Now substitute these expressions into equations \ref{field2}, then the results into equations \ref{field1},
multiply each resulting equation by $e^{-i2\pi({\frac{n_1x_1}{a_1}+\frac{n_2x_2}{a_2}})}$ and integrate over the unit cell obtain,
\begin{eqnarray}\label{p6}
 \begin{array}{c}
H_1S_{11}+H_2S_{12}+\Lambda_{\rho}\omega^2U_{1}=0\\
H_1S_{21}+H_2S_{22}+\Lambda_{\rho}\omega^2U_{2}=0\\
\end{array} , ~~~
\left[ \begin{array}{c}
H_1U_1\\
H_2U_2\\
H_1U_2+H_2U_1\\
\end{array} \right]=
\left[ \begin{array}{c}
\Lambda_{D_{11}}S_{11}+\Lambda_{D_{12}}S_{22}\\
\Lambda_{D_{22}}S_{22}+\Lambda_{D_{12}}S_{11}\\
\Lambda_DS_{12}\\
\end{array}\right],
\end{eqnarray} 
where 
$S_{jl}=[S_{jl}^{n_1,n_2}]$ and
$U_j=[U_j^{n_1,n_2}]$ are each  
a $(2N+1)^2 \times(2N+1)^2$ matrix,
$H_1,H_2$ are $(2N+1)^2 \times(2N+1)^2$  diagonal matrices with respective diagonal components 
i($k_1+2\pi \frac{n_1}{a_1})\delta_{n_1m_1}$, and
i($k_2+2\pi \frac{n_2}{a_2})\delta_{n_2m_2}$.
The components of $\Lambda_f$, with $f$ standing for any of the material parameters, are defined by
\begin{equation}\label{p7}%14
\Lambda^{(n_1n_2,m_1m_2)}_{f}=
 \int_{\Omega} f(\mathbf{x}) 
e^{i2\pi 
[\frac{(n_1 - m_1)x_1}{a_1}+
\frac{(n_2 - m_2)x_2}{a_2}]}
dx_1 dx_2,
\end{equation}
and $\Lambda_D= \Lambda_{D_{11}}+ \Lambda_{D_{22}}-2 \Lambda_{D_{12}}$.
Eliminating $S_{12}$ in equations (\ref{p6}), we obtain,
\begin{eqnarray}\label{p8}%15
\hspace*{-2cm}
\left[ \begin{array}{cc}
\Lambda&-H\\
H&\Psi\\
\end{array} \right]\
\left[ \begin{array}{c}
S\\
U\\
\end{array} \right]=0,\quad
\Lambda=
\left[ \begin{array}{cc}
\Lambda_{D_{11}} & \Lambda_{D_{12}} \\
\Lambda_{D_{21}} & \Lambda_{D_{22}} \\
\end{array} \right],\quad
H=\left[ \begin{array}{cc}
H_1&0\\
0&H_2\\
\end{array} \right],
\end{eqnarray}    
\begin{eqnarray}\label{p9}%16
%\hspace*{-2cm}
\Psi=
\left[\begin{array}{ccc}
H_2\Lambda_D^{-1}H_2 +\omega^2\Lambda_{\rho}&H_2\Lambda_D^{-1}H_1\\
H_1\Lambda_D^{-1}H_2&H_1\Lambda_D^{-1}H_1 +\omega^2\Lambda_{\rho}\\
\end{array} \right],\quad
S=
\left[\begin{array}{c}
S_{11}\\
S_{22}\\
\end{array} \right],\quad
U=
\left[\begin{array}{c}
U_1\\
U_2\\
\end{array} \right].
\end{eqnarray}  
Finally, eliminating the stresses, $S$, in favor of the displacements, we arrive at the following eigenvalue problem:
\begin{equation}\label{p10}%17
[
H\Lambda^{-1}H+\Psi+\omega^2\Lambda_{\rho}I
]
U=0,\quad
det |H\Lambda^{-1}H+\Psi+\omega^2\Lambda_{\rho}I|=0,
\end{equation}
where $I$ is the identity matrix. The roots of equation (\ref{p10})$_2$ are 
the frequency bands, $\omega$, defined as functions of  $Q_1$ and $Q_2$. For each eigenfrequency, $U_j$'s  are obtained from
equation (\ref{p10})$_1$, and then the corresponding  stresses are given by
$
S=\Lambda^{-1}HU,
$
and 
$
S_{12}=S_{21}=\Lambda_D^{-1}(H_1U_2+H_2U_1)
$.

\textit{Group Velocity and Energy Flux}:
The determinant in equation  (\ref{p10})$_2$ depends parametrically on the wave-vector components, $Q_1\equiv k_1a_1$ and $Q_2\equiv k_2a_2$.   The resulting eigenfrequencies, $\omega$, are thus functions of $Q_1$ and $Q_2$. These eigenfrequencies form surfaces in the ($Q_1$, $Q_2$,  $\omega$)-space, referred to as the Brillouin zones. The  fundamental zone corresponds to $-\pi{\leq{Q_1,Q_2}\leq }\pi$. Our focus in the present work has been on the acoustic branch of this zone, which includes two passband, one shear and the other longitudinal mode.
On each of these frequency bands, the phase and group velocities are given by
\begin{equation}\label{ph-group}
v^p_{Jj}=\frac{\omega_Jk_j}{k_1^2+k_2^2},\quad
v^g_{Jj}=\frac{\partial \omega_J}{\partial{k_j}},\quad
j=1,2;
\end{equation}
here and below, $J=1, 2$ denotes the shear and longitudinal frequency band,respectively, and $j=1,2$.   
The refraction angle, say $\alpha_J$, is computed from
\begin{equation}\label{Refraction_Group}
\alpha_{J}=atan(\frac{v^g_{J2}}{v^g_{J1}}).
\end{equation}
It is known that the direction, $\alpha_J$, is essentially the same as the direction of the energy flux for non-dissipative media; for a general proof, see\cite{willis2016negative}.  Below, the calculation of the energy flux and its direction direction are outlined.

\begin{figure}[htbp]
\begin{center}
\begin{minipage}[b]{0.45\linewidth}
\includegraphics[scale=0.40, trim=0.cm 0.cm 0.cm 0.cm, clip=true]{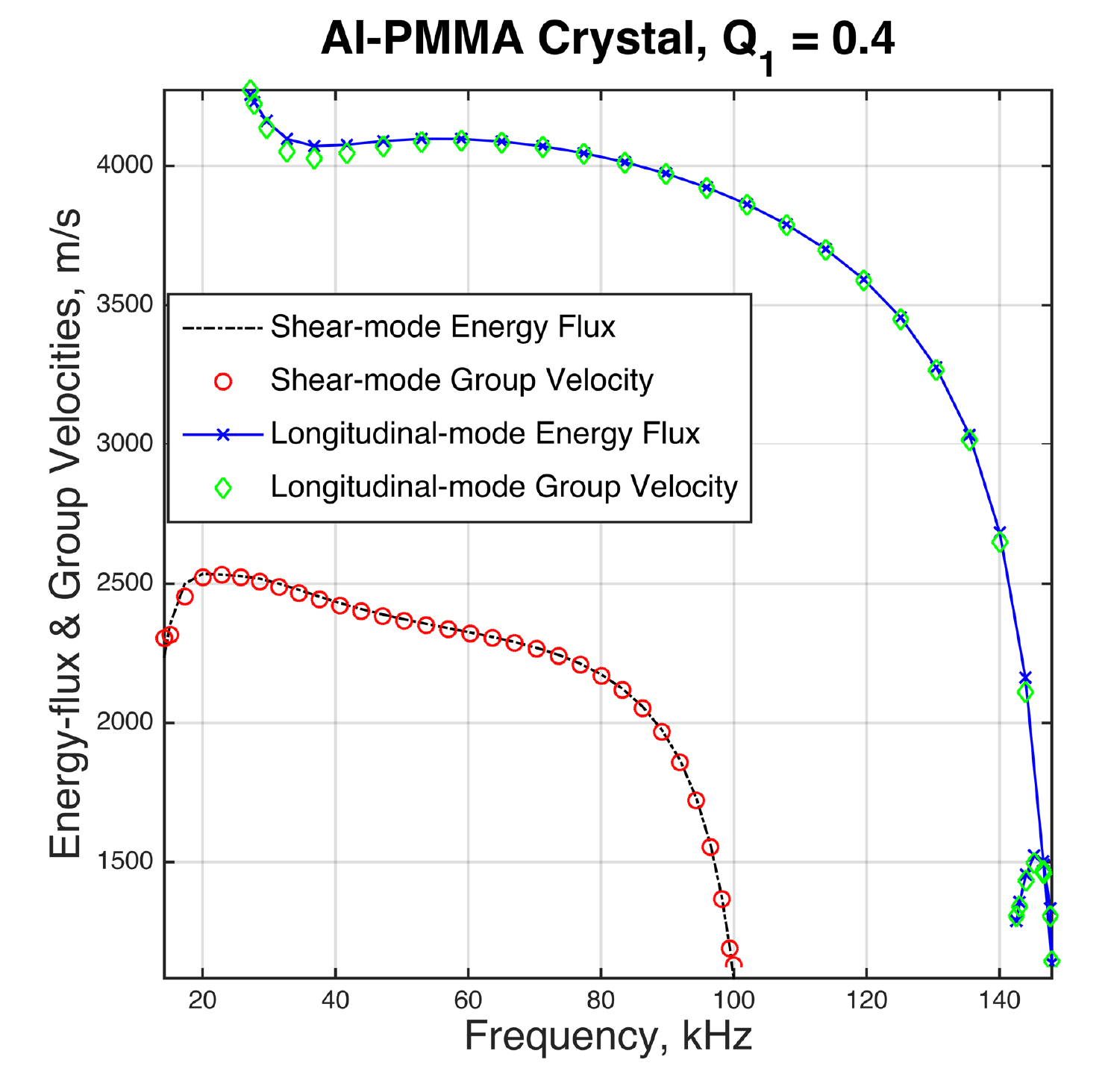}
\end{minipage}
\begin{minipage}[b]{0.45\linewidth}
\includegraphics[scale=0.40, trim=0.cm 0.cm 0.cm 0.cm, clip=true]{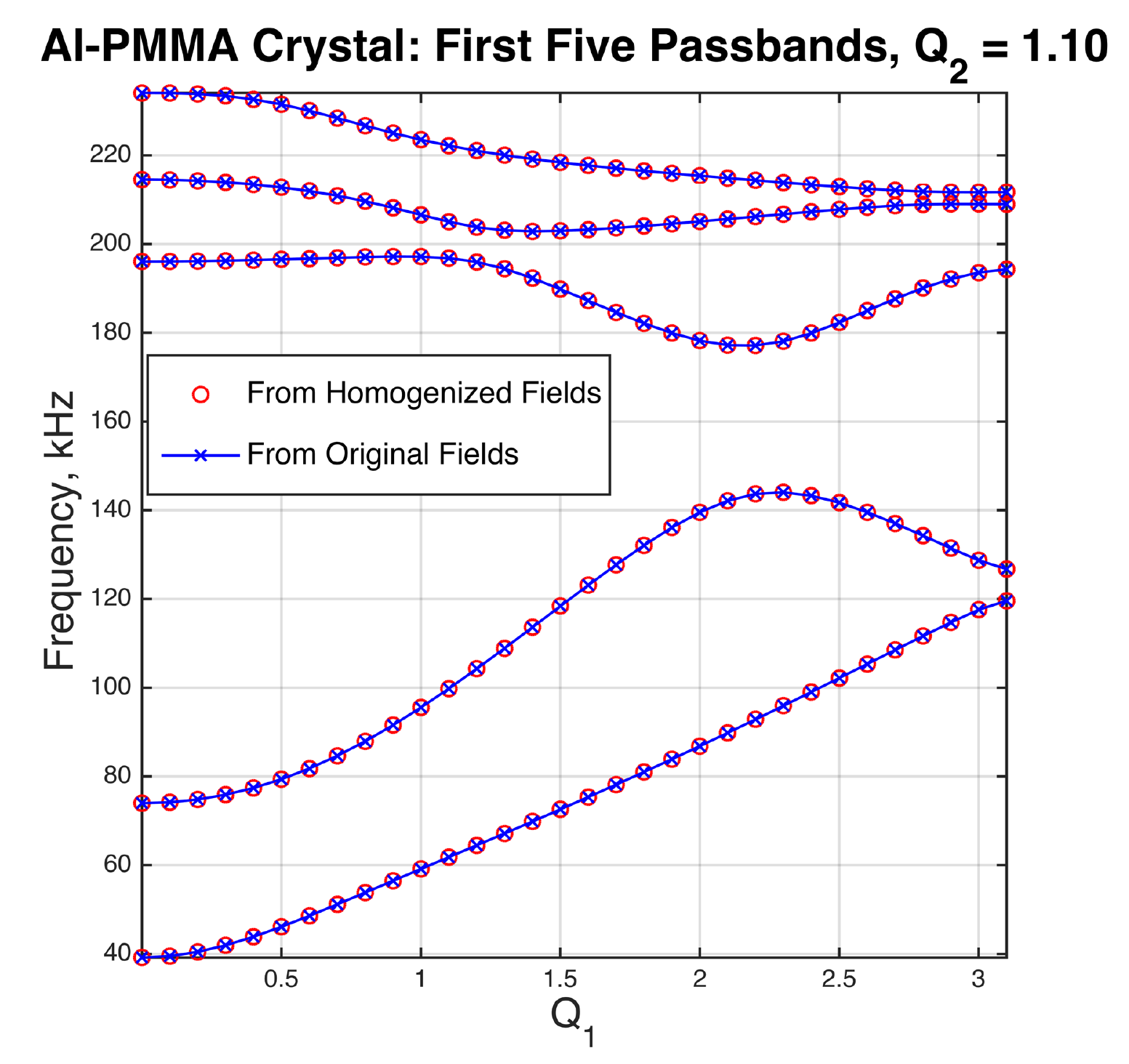}
\end{minipage}
\caption{(Left): Acoustic-branch shear and longitudinal energy-flux and group velocities as function of frequency for $Q_1=0.4$; (Right) First passbands frequencies for $Q_2=1.10$.}
\label{Bands}
\end{center}
\end{figure}

The $x_1$- and $x_2$-components of the  energy flux, averaged over a unit cell, are given by
\begin{equation} \label{Energy-flux}
\bar{E}_{Jk}  =
\frac{\omega_J}{2\pi}
\int_{0}^{2\pi/\omega_J}
\hspace{-0.5cm}
< Re(\sigma_{kjJ})
Re(\dot{u}_{jJ})^* > dt
=-\frac{1}{2}<\sigma_{kjJ}^p\dot{u}_{jJ}^{p*}>\\
 = \frac{1}{2}i\omega_J
 \hspace{-0.5cm}
 \sum_{n_1,n_2=-N}^{+N}S_{kjJ}^{n_1,n_2}U_{jJ}^{n_1,n_2},
\end{equation} 
where $k,j=1,2$. 
The direction, $ {\beta_J} $,  of the energy-flux vector is hence given by,
\begin{equation}\label{Refraction_Energy}
\beta_J=atan(\frac{\bar{E}_{J2}}{\bar{E}_{J1}}).
\end{equation}
It turns out that $\alpha_J=\beta_J$ for the class of problems considered in the present work,
as has been demonstrated in Figures \ref{expS1}(Left) and \ref{expL1}(Left). 
The components of the energy-flux velocity is readily obtained by dividing the flux components, $\bar{E}_{Jk}$, by the corresponding average elastic strain energy, 
\begin{equation}\label{Energy-fluxVector}
\bar{V}_{Jk}=\frac{\bar{E}_{Jk}}{\frac{1}{2}<\sigma_{ljJ}^p\epsilon_{ljJ}^{p*}>}.
\end{equation}
Here again the group velocity and the energy-flux velocity are essentially identical, as illustrated in Figure \ref{Bands}(Left).

\textit{Homogenization}:
The nonzero components of the effective compliance and  effective mass-density matrixes are defined as follows:
\begin{align}\label{eff} \nonumber
&D_{11}^{eff}=\frac{\overline{(D_{11}\sigma_{11}^p)}}{\bar{\sigma}_{11}^p},
D_{12}^{eff}=\frac{\overline{(D_{12}\sigma_{22}^p)}}{\bar{\sigma}_{22}^p},
D_{21}^{eff}=\frac{\overline{(D_{21}\sigma_{11}^p)}}{\bar{\sigma}_{11}^p},
D_{22}^{eff}=\frac{\overline{(D_{22}\sigma_{22}^p)}}{\bar{\sigma}_{22}^p},
D_{33}^{eff}=\frac{\overline{(D_{33}\sigma_{12}^p)}}{\bar{\sigma}_{12}^p},\\
&\rho_1^{eff}=\frac{\overline{(\rho u_1^p)}}{\bar{u}_1^p},~~
\rho_2^{eff}=\frac{\overline{(\rho u_2^p)}}{\bar{u}_2^p},~
\end{align}

where superimposed bar stands for integration over a unit cell.
The homogenized field equations are obtained by integrating the periodic part of the corresponding equations, arriving at,
\begin{equation}
ik_l\bar{\sigma}_{jl}^p+\omega^2\rho_j^{eff}\bar{u}_j=0,~~~
[\bar{\epsilon}^p]=[D^{eff}][\bar{\sigma}^p],~~~
\bar{\epsilon}_{jl}^p=\frac{i}{2}(k_j\bar{u}_l^p+k_l\bar{u}_j^p),~~~j,l=1,2.
\end{equation} 
Combining the above equations and setting,
\begin{align}\label{eff} \nonumber
&[\Psi^{eff}]=
\left[\begin{array}{cc}
-k_2^2\mu^{eff}-k_1^2\mu_{11}^{eff}+\omega^2\rho_1^{eff}&
-k_1k_2(\mu^{eff}+\mu_{12}^{eff})\\
-k_1k_2(\mu^{eff}+\mu_{21}^{eff})&
-k_1^2\mu^{eff}-k_2^2\mu_{22}^{eff}+\omega^2\rho_2^{eff}
\end{array}\right],\\
&[\mu^{eff}]=
\left[ \begin{array}{cc}
D_{11}^{eff}&D_{12}^{eff}\\
D_{21}^{eff}&D_{22}^{eff}
\end{array}\right]^{-1},
\end{align}
we obtian,
\begin{equation}\label{om}
[\Psi^{eff}][\bar{u}]=0,~~~
det|\Psi^{eff}|=0,
\end{equation}
where $2\mu^{eff}=(D_{33}^{eff})^{-1}$, and the eigenvalues are the roots of equation (\ref{om}).

\textit{Acknowledgement}:
This research has been conducted at the Center of Excellence for Advanced Materials (CEAM) at the University of California, San Diego. The author wishes to thank Mr. Shailendra  Singh for his assistance with the sample preparation and data acquisition.  

%\section*{References}

%\bibliography{REFS_Harmonic-1.bib}

\begin{thebibliography}{10}
\expandafter\ifx\csname url\endcsname\relax
  \def\url#1{\texttt{#1}}\fi
\expandafter\ifx\csname urlprefix\endcsname\relax\def\urlprefix{URL }\fi
\expandafter\ifx\csname href\endcsname\relax
  \def\href#1#2{#2} \def\path#1{#1}\fi

\bibitem{Smith2000}
D.~R. Smith, W.~J. Padilla, D.~Vier, S.~C. Nemat-Nasser, S.~Schultz, Composite
  medium with simultaneously negative permeability and permittivity, Physical
  review letters 84~(18) (2000) 4184.

\bibitem{shelby2001experimental}
R.~A. Shelby, D.~R. Smith, S.~Schultz, Experimental verification of a negative
  index of refraction, Science 292~(5514) (2001) 77--79.

\bibitem{minagawa1976harmonic}
S.~Minagawa, S.~Nemat-Nasser, Harmonic waves in three-dimensional elastic
  composites, International Journal of Solids and Structures 12~(11) (1976)
  769--777.

\bibitem{nemat2015refraction}
S.~Nemat-Nasser, Refraction characteristics of phononic crystals, Acta
  Mechanica Sinica 31~(4) (2015) 481--493.

\bibitem{nemat2017unified}
S.~Nemat-Nasser, Unified homogenization of photonic/phononic crystals with
  first-band negative refraction, Mechanics of Materials 105 (2017) 29--41.

\bibitem{yang2004focusing}
S.~Yang, J.~Page, Z.~Liu, M.~Cowan, C.~Chan, P.~Sheng, Focusing of sound in a
  3d phononic crystal, Physical review letters 93~(2) (2004) 024301.

\bibitem{ke2005negative}
M.~Ke, Z.~Liu, C.~Qiu, W.~Wang, J.~Shi, W.~Wen, P.~Sheng, Negative-refraction
  imaging with two-dimensional phononic crystals, Physical Review B 72~(6)
  (2005) 064306.

\bibitem{sukhovich2008negative}
A.~Sukhovich, L.~Jing, J.~H. Page, Negative refraction and focusing of
  ultrasound in two-dimensional phononic crystals, Physical Review B 77~(1)
  (2008) 014301.

\bibitem{notomi2000theory}
M.~Notomi, Theory of light propagation in strongly modulated photonic
  crystals: Refractionlike behavior in the vicinity of the photonic band gap,
  Physical Review B 62~(16) (2000) 10696.

\bibitem{Luo2002all}
C.~Luo, S.~G. Johnson, J.~Joannopoulos, J.~Pendry, All-angle negative
  refraction without negative effective index, Physical Review B 65~(20) (2002)
  201104.

\bibitem{willis2016negative}
J.~Willis, Negative refraction in a laminate, Journal of the Mechanics and
  Physics of Solids 97 (2016) 10--18.

\end{thebibliography}

\end{document}